\documentclass[copyright,creativecommons,noderivs,noncommercial]{eptcs}
\providecommand{\event}{WLP 2016} 
\usepackage{breakurl}             

\makeatletter

\def\define#1{\@ifnextchar [{\@MYargdef#1}{\@MYargdef#1[0]}}

\def\@MYargdef#1[#2]#3{\@ifdefinable #1{\@MYreargdef#1[#2]{#3}}}

\def\redefine#1{\edef\@tempa{\expandafter\@cdr\string 
  #1\@nil}\@ifundefined{\@tempa}{\@latexerr{\string#1\space undefined}\@ehc
    }{}\@ifnextchar [{\@MYreargdef#1}{\@MYreargdef#1[0]}}

\catcode`\?=11\relax
\def\@MYreargdef#1[#2]#3{\@tempcnta#2\relax\let#1\relax
\edef\@tempa{\def#1}\@tempcntb \@ne
\let\@?@?\relax\@whilenum\@tempcnta>0
\do{\edef\@tempa{\@tempa\@?@?\the\@tempcntb}\advance\@tempcntb \@ne \advance
\@tempcnta \m@ne}\let\@?@?##\@tempa{#3}}
\catcode`\?=12\relax

\makeatother

\renewcommand{\bmod}%
{\mskip-\medmuskip \mkern5mu \mathbin{\idFont mod} \mkern5mu \mskip-\medmuskip}

\define{\mathsym}[1]{\relax\ifmmode#1\else
	\errmessage{Mathematical symbol outside math mode}\fi}
\define{\idFont}{\sf}
\define{\id}[1]{\mathsym{{\idFont #1}}}

\define{\m}[1]{$#1$}
\define{\M}[1]{$$#1$$}

\define{\gt}{\mathsym{\mathchar 12606\relax}} 
\define{\lt}{\mathsym{\mathchar 12604\relax}} 

\define{\paren}[1]{(#1)}
\define{\parenA}[1]{(#1)}
\define{\parenB}[1]{\bigl(#1\bigr)}
\define{\parenC}[1]{\Bigl(#1\Bigr)}
\define{\parenAuto}[1]{\left(#1\right)}
\define{\set}[1]{\lbrace#1\rbrace}
\define{\setA}[1]{\lbrace#1\rbrace}
\define{\setB}[1]{\bigl\lbrace#1\bigr\rbrace}
\define{\setC}[1]{\Bigl\lbrace#1\Bigr\rbrace}
\define{\setCond}[2]{\lbrace#1\mathrel{\vert}#2\rbrace}
\define{\setCondA}[2]{\lbrace#1\mathrel{\vert}#2\rbrace}
\define{\setCondB}[2]{\bigl\lbrace#1\bigm\vert#2\bigr\rbrace}
\define{\setCondC}[2]{\Bigl\lbrace#1\Bigm\vert#2\Bigr\rbrace}

\define{\union}{\cup}
\define{\unionUntil}{\union\cdots\union}
\define{\unionMulti}[2]{\bigcup_{#1}^{#2}}
\define{\intersect}{\cap}
\define{\intersectUntil}{\intersect\cdots\intersect}
\define{\intersectMulti}[2]{\bigcap_{#1}^{#2}}
\define{\timesUntil}{\times\cdots\times}
\define{\setsize}[1]{{\left\vert #1 \right\vert}}
\define{\compl}[1]{\overline{#1}}

\define{\powerset}[1]{2^{#1}}
\define{\inB}{\;\in\;}

\define{\defEq}{:=}
\define{\defEqB}{\;:=\;}
\define{\defIff}{\;\mathsym{:\Longleftrightarrow}\;}
\define{\metaThen}{\mathsym{\;\Longrightarrow\;}}
\define{\metaIf}{\mathsym{\;\Longleftarrow\;}}
\define{\metaIff}{\mathsym{\;\Longleftrightarrow\;}}

\define{\lneg}{\mathsym{\neg}}
\define{\lif}{\mathsym{\leftarrow}}
\define{\lifDup}{\mathsym{\Leftarrow}}
\define{\lthen}{\mathsym{\rightarrow}}
\define{\liff}{\mathsym{\leftrightarrow}}
\define{\lfalse}{\id{false}}
\define{\ltrue}{\id{true}}

\define{\lorUntil}{\lor\cdots\lor}
\define{\landUntil}{\land\cdots\land}
\define{\lorMulti}[2]{\bigvee_{#1}^{#2}}
\define{\landMulti}[2]{\bigwedge_{#1}^{#2}}

\define{\lfor}[4]{#1\:#2\:#3:#4}
\define{\lexistsQ}{\exists}
\define{\lexists}[3]{\lfor{\lexistsQ}{#2}{#1}{#3}}
\define{\lallQ}{\forall}
\define{\lall}[3]{\lfor{\lallQ}{#2}{#1}{#3}}

\define{\answerPred}{\id{answer}}

\define{\natNum}{\mathsym{{\rm l\kern-0.13em N}}}

\define{\realNum}{\mathsym{{\rm I\kern-0.14em R}}}

\define{\struct}[1]{\langle#1\rangle}
\define{\twoCases}[4]{\left\{\begin{array}{l@{\kern10pt}l}
				#1&\mbox{#2}\\#3&\mbox{#4}
				\end{array}\right.}
\define{\until}{, \ldots,}
\define{\seqOf}[1]{#1^*}
\define{\emptySeq}{\epsilon}
\define{\w}{\id{w}}

\define{\code}[1]{{\tt #1}} 

{\begin{center}\begin{tt}\begin{tabular}{@{}l@{}}}%
{\end{tabular}\end{tt}\end{center}}

	{\begin{center}\begin{tt}\begin{tabular}{@{}l@{ }l@{}}}%
	{\end{tabular}\end{tt}\end{center}}

\define{\U}{{\char95}} 
\define{\LT}{{\char60}} 
\define{\GT}{{\char62}} 
\define{\B}{{\char92}} 
\define{\AMP}{{\char38}} 
\define{\D}{{\char36}} 
\define{\SN}{{\char126}} 
\define{\HASH}{{\char35}} 
\define{\Q}{{\char34}} 
\define{\PCT}{{\char37}} 
\define{\LB}{{\char123}} 
\define{\RB}{{\char125}} 
\define{\HAT}{{\char94}} 

\define{\ALPH}{\id{ALP\kern-0.08em H}}
\define{\LOG}{\id{LOG}}
\define{\VARS}{\id{V\kern-0.17em ARS}}
\define{\var}{\mathsym{X}}
\define{\varA}{\id{X}}
\define{\varB}{\id{Y}}
\define{\varC}{\id{Z}}
\define{\const}{\mathsym{c}}
\define{\constA}{\id{a}}
\define{\constB}{\id{b}}
\define{\constC}{\id{c}}
\define{\data}{\mathsym{d}}
\define{\dataSet}{\mathsym{{\cal D}}}
\define{\sig}{\Sigma}
\define{\SORTS}{\id{{\cal S}}}
\define{\sort}{\id{s}}
\define{\PREDS}{\id{{\cal P}}}
\define{\pred}{\id{p}}
\define{\predA}{\id{p}}
\define{\predB}{\id{q}}
\define{\predC}{\id{r}}
\define{\predD}{\id{s}}
\define{\arity}{\id{n}}
\define{\level}{l}
\define{\FUNS}{\id{{\cal F}}}
\define{\fun}{\id{f}}
\define{\args}{\alpha}
\define{\argsB}{\beta}
\define{\argSorts}{\alpha}
\define{\resSort}{\rho}
\define{\interp}{\id{{\cal I}}}
\define{\interpB}{\id{{\cal J}}}
\define{\ass}{\id{{\cal A}}}
\define{\iV}[2]{(#1,#2)}
\define{\eval}[2]{#1\lbrack\kern-0.15em\lbrack#2\rbrack\kern-0.15em\rbrack}
\define{\evalV}[3]{\eval{\iV{#1}{#2}}{#3}}
\define{\varDecl}{\nu}
\define{\TERMS}{\id{T\kern-0.1em E}}
\define{\term}{\id{t}}
\define{\termB}{\id{u}}
\define{\termC}{\id{v}}
\define{\argA}{\id{a}}
\define{\argB}{\id{b}}
\define{\argC}{\id{c}}
\define{\AT}{\id{AT}}
\define{\FO}{\id{FO}}
\define{\fo}{\varphi}
\define{\foB}{\psi}
\define{\fos}{\Phi}
\define{\modify}[3]{#1\langle#2/#3\rangle}
\define{\impl}{\vdash}
\define{\subst}{\theta}
\define{\substB}{\sigma}
\define{\mgu}{\id{mgu}}
\define{\doSubst}[2]{#2#1}
\define{\doSubstB}[2]{(#2)\,#1}
\define{\HU}{\id{{\cal U}}}
\define{\HB}{\id{{\cal B}}}
\define{\HSet}{\id{H}}
\define{\fact}{\id{F}}
\define{\ground}{\id{ground}}
\define{\LAT}{\id{{\cal M}}}
\define{\LATB}{\id{{\cal N}}}

\define{\emptyClause}{{\hbox{%
	\setlength{\unitlength}{0.24ex}%
	\begin{picture}(5,5)(0,0)
	\put(0,0){\line(0,1){5}}
	\put(0,0){\line(1,0){5}}
	\put(0,5){\line(1,0){5}}
	\put(5,0){\line(0,1){5}}
	\end{picture}}}}

\define{\lit}{\id{L}}
\define{\litA}{\id{A}}
\define{\litB}{\id{B}}
\define{\litC}{\id{C}}
\define{\Body}{\mathsym{{\cal B}}}
\define{\F}{\id{F}} 
\define{\ruleFun}[1]{\id{r}_{#1}}

\define{\prog}{\mathsym{{\idFont P}}}
\define{\T}[1]{\mathsym{{\idFont T}}_{#1}}
\define{\TP}{\T{\prog}}
\define{\Tneg}[2]{\mathsym{{\idFont T}}_{#1,#2}}
\define{\lub}{\id{lub}}
\define{\glb}{\id{glb}}
\define{\lfp}{\id{l\kern-0.1em f\kern-0.1em p}}

\define{\I}{\id{{\cal I}}}
\define{\J}{\id{{\cal J}}}

\define{\nf}{\mathop{\id{not\kern0.2em}}}
\define{\nfPred}[1]{\mathop{\id{not{\U}}#1}}

\define{\free}{\id{f}}
\define{\bound}{\id{b}}

\define{\bp}{\id{bp}}
\define{\binding}{\beta}
\define{\bindingSet}{{\cal B}}

\define{\vars}{\mathsym{{\cal X}}}
\define{\varsA}{\mathsym{{\cal X}}}
\define{\varsB}{\mathsym{{\cal Y}}}
\define{\varsC}{\mathsym{{\cal Z}}}

\define{\inputVars}{\id{input}}

\define{\freeVar}{\id{vars}}
\define{\varsOf}{\id{vars}}
\define{\boundPos}{\id{bound}^+}
\define{\boundNeg}{\id{bound}^-}
\define{\unboundPos}{\id{unbound}^+}
\define{\unboundNeg}{\id{unbound}^-}
\define{\err}{\id{err}}

\newcounter{ProgramLine}
\newcounter{FirstLine}

\newenvironment{progTabular}{%
	\addtocounter{ProgramLine}{-1}%
	\setcounter{FirstLine}{1}%
	\ifvmode\vspace{5mm}\fi%
	\begin{list}{\(\bullet\)\hfill}{
		\parskip 3pt plus 1pt
		\labelwidth 0pt
		\labelsep 0pt
		\leftmargin 8mm
		\listparindent \parindent
		\topsep 4mm
		\parsep 3pt plus 1pt
		\itemsep 0pt
		\partopsep 0pt
		\itemindent 0pt
		\rightmargin 0pt
	}
	\item[]
	\begin{tabular}{@{}r@{\hspace{2mm}}l@{}}}%
	{\end{tabular}%
		\end{list}}

\newenvironment{progPart}[1]{%
		\setcounter{ProgramLine}{#1}%
		\begin{progTabular}}%
	{\end{progTabular}}

	{\end{progTabular}%
		\end{tt}}

	{\end{progTabular}%
		\end{tt}}

\define{\tabX}[1]{\ifnum\value{FirstLine}=1\setcounter{FirstLine}{0}\else
	\ifhmode\\[0.5pt]\fi\fi
	\stepcounter{ProgramLine}(\arabic{ProgramLine})&
	\hspace*{#1}}
\define{\tabA}{\tabX{0cm}}
\define{\tabB}{\tabX{8mm}}
\define{\tabC}{\tabX{16mm}}
\define{\tabD}{\tabX{24mm}}
\define{\tabE}{\tabX{32mm}}
\define{\tabF}{\tabX{40mm}}
\define{\tabG}{\tabX{50mm}}
\define{\tabH}{\tabX{60mm}}
\define{\tabBox}[1]{\vrule height0pt depth0pt width0pt\hbox to14mm{#1\ \hfil}}

\define{\IF}{{\bf if }}
\define{\THEN}{{\bf then }}
\define{\ELSE}{{\bf else }}
\define{\FI}{{\bf fi}}
\define{\FOREACH}{{\bf foreach }}
\define{\FOR}{{\bf for }}
\define{\TO}{{\bf to }}
\define{\FROM}{{\bf from }}
\define{\IN}{{\bf in }}
\define{\WITH}{{\bf with }}
\define{\AND}{{\bf and }}
\define{\NOT}{{\bf not }}
\define{\TRUE}{{\bf true}}
\define{\FALSE}{{\bf false}}
\define{\DO}{{\bf do }}
\define{\OD}{{\bf od}}
\define{\WHILE}{{\bf while }}
\define{\BREAK}{{\bf break}}
\define{\PROCEDURE}{{\bf procedure }}
\define{\BEGIN}{{\bf \code{\LB}} }
\define{\END}{{\bf \code{\RB}} }
\define{\RETURN}{{\bf return }}
\define{\LET}{{\bf let }}
\define{\NEW}{{\bf new }}
\define{\COMPUTE}{{\bf compute }}
\define{\APPEND}{{\bf append }}
\define{\PRINT}{{\bf print }}
\define{\BOOL}{{\bf bool }}
\define{\NIL}{{\bf nil }}
\define{\OUTPUT}{{\bf output }}
\define{\INSERT}{{\bf insert }}
\define{\INTO}{{\bf into }}
\define{\CHOOSE}{{\bf choose }}

\define{\COMMENT}[1]{{\rm /\raisebox{-.6ex}{*} #1 \raisebox{-.6ex}{*}/}}

\define{\CPP}{C{\tt ++}}

\define{\activePart}[1]{\id{a}(#1)}
\define{\delayedPart}[1]{\id{d}(#1)}

\define{\lineno}[1]{\hbox to 1.6em{\hfil\m{\lbrack#1\rbrack}}}
\define{\lineref}[1]{\m{\lbrack#1\rbrack}}
\define{\comment}[1]{\mbox{// #1}}

\define{\C}{\mathsym{C}}

\define{\ruleNo}{\rho}

\define{\edbPredA}{\id{e}}
\define{\edbPredB}{\id{r}}
\define{\edbPredC}{\id{s}}

\define{\state}{\mathsym{{\cal S}}}
\define{\Rule}{\mathsym{{R}}}
\define{\db}{\mathsym{{\cal D}}}

\define{\vertices}{\mathsym{{\cal V}}}
\define{\edges}{\mathsym{{\cal E}}}

\define{\called}[2]{(\lineref{#1},\lineref{#2})}

\define{\grandparent}{\id{grandparent}}
\define{\parent}{\id{parent}}
\define{\father}{\id{father}}
\define{\mother}{\id{mother}}
\define{\personA}{\id{ann}}
\define{\personB}{\id{betty}}
\define{\personC}{\id{chris}}
\define{\personD}{\id{david}}

\newtheorem{example}{Example}
\define{\qed}{\hfill\fbox{\phantom{\rule{.65ex}{.65ex}}}}

\title{Experiences with Some Benchmarks for Deductive Databases
	and Implementations of Bottom-Up Evaluation}

\author{Stefan Brass\quad\quad Heike Stephan
	\institute{Institut f\"ur Informatik,
		Martin-Luther-Universit\"at Halle-Wittenberg,
		Germany}\\
	\email{brass@informatik.uni-halle.de
		\quad\quad stephan@informatik.uni-halle.de}
}

\def\titlerunning{Experiences with Benchmarks for Deductive Databases}
\def\authorrunning{S.~Brass \& H. Stephan}

\begin{document}

\maketitle



\begin{abstract}
OpenRuleBench~\cite{LFWK09} is a large benchmark suite
for rule engines,
which includes deductive databases.
We previously proposed
a translation of Datalog
to \CPP\ based on a method that ``pushes''
derived tuples immediately to places where they are used.
In this paper,
we report performance results
of various implementation variants of this method
compared to XSB, YAP and DLV.
We study only a fraction of the OpenRuleBench problems,
but we give a quite detailed analysis of each such task
and the factors which influence performance.
The results not only show the potential of our method
and implementation approach,
but could be valuable for anybody
implementing systems which should be able to execute
tasks of the discussed types.
\end{abstract}


 


\section{Introduction}
\label{secIntro}

With deductive database technology,
the range of tasks a database can do is increasing:
It can not only fetch and compose data,
as classical SQL queries,
but a larger part of the application can be specified declaratively
in a pure logic programming language.
It is common that database systems offer stored procedures
and triggers today,
but this program code is written in an imperative language,
not declaratively as queries in SQL.
In addition, still other (non-declarative) languages are used
for the application development itself.
It is the claim of deductive databases like LogicBlox~\cite{Aref15}
to unify all this.
%

Making deductive databases a successful programming platform
for future applications has many aspects,
including the design of language features
for declarative, rule-based output
(see, e.g.,~\cite{Bra12}).
But improving the performance of query evaluation / program execution
is still an important task.
Some older deductive database prototypes
have been not very good from the performance viewpoint.
Typically, people who are not yet convinced of the declarative approach		
ask about performance as one of their first questions.
Furthermore, the amount of data to be processed
is constantly increasing.
New processor architectures offer new opportunities,
which may not be utilized in old code.

Benchmarks can be a useful tool to evaluate and compare the performance
of systems.
Furthermore,
they can be motivating for new system developers.
In the area of deductive databases,
OpenRuleBench~\cite{LFWK09} is a well-known benchmark suite.
It is a quite large collection of problems,
basically 12~logic programs,
but some with different queries and different data files.
The original paper~\cite{LFWK09} contains 18~tables with benchmark results.
The logic programs used for the tests
(with different syntax variants and settings for different systems),
the data files and supporting shell scripts can be downloaded from
\begin{center}
\url{http://www3.cs.stonybrook.edu/~pfodor/openrulebench/download.html}
\end{center}
The tests have been re-run in 2010 and 2011,
the results are available at:
\begin{center}
\url{http://rulebench.semwebcentral.org/}
\end{center}
In this paper,
we look at only five of the benchmark problems,
but in significantly more detail
than the original OpenRuleBench paper.
We are currently developing an implementation
of our ``Push''-method for bottom-up evaluation~\cite{Bra10CPP,BS15WLP},
and would like to document and to share our insights
from many preformance tests with different variants.
We compare our method with XSB~\cite{SSW94},
YAP~\cite{CRD12},
and dlv~\cite{LPFEGPS06}.
It is also of some value that we did these measurements
again five years later with new versions of these systems
on a different hardware.
We also put more emphasis on the loading time and overall runtime
than the authors of the original OpenRuleBench paper~\cite{LFWK09}.
In addition,
we did some measurements on main memory consumption.

As we understand the rules of OpenRuleBench,
the options for each benchmark and system
were manually selected in cooperation with the system developers.
This means that a good automatic optimization
does not ``pay off'' in the benchmark results,
although it obviously is of importance for the usability of the system.
However,
this also helps to separate two different aspects:
The basic performance of program execution
and having an intelligent optimizer,
which finds a good query execution plan
and index structures.
For systems that allow to influence query evaluation
with many options,
the optimizer becomes less important for the OpenRuleBench results.

Since our query execution code
is in part manually created
and we still try different data structures,
this can be seen as an extreme case of the manual tuning.
It is our goal to test and demonstrate the potential
of our approach,
but we are still far from a complete system.
Naturally,
before investing maybe years to develop a real deductive database system,
it is good to see what can be reached,
and whether it is worth to follow this path.

In contrast,
each of the three systems we use as a comparison	
has been developed over more than a decade.
They are mature systems which offer a lot of programming features
which are important for usage in practical projects.
Although our performance numbers are often significantly better,
our prototype could certainly not be considered
as a competitor for real applications.

Our approach is to translate Datalog (pure Prolog without function symbols)
to \CPP,
which is then compiled to native code
by a standard compiler.
We have an implementation of this translation (written in Prolog),
and a library of data structures used in the translated code
written in \CPP:
\begin{center}
\url{http://www.informatik.uni-halle.de/~brass/push/}
\end{center}
At the moment,
the result of the translation must still be manually copied
to a \CPP\ class frame,
and compiler and linker are manually executed,
but it should be an easy task to automize this.
The tests reported in the current paper discuss
different code structures,
therefore it seemed better to first find the optimal structure
before programming the complete translation.
However,
the current implementation of the Datalog-to-\CPP\ translation
already generates different variants of the main query execution code.

XSB compiles into machine code of an abstract machine
(XWAM) and interprets that code.
The same is true for YAP~\cite{CRD12}
(the abstract machine is called YAAM).
DLV probably uses a fully interpreted approach.
Since we compile to native code,
this obviously gives us an advantage.
If one compares the numbers given in~\cite{Costa99}
for Sicstus Prolog native code vs.~emulated bytecode,
the speedup for native code is between 1.5 and~5.6,
with median value~2.6.
%
Of course,
the speedup also depends on the granularity of the operations
that are executed.
For instance,
if many operations are index lookups or other large database operations,
these are in native code in the virtual machine emulator anyway,
so that the overhead of interpretation is not important in this case.
But this means that the speedup that must be attributed
to using native code might even be less than what is mentioned above.


One of the strengths of the ``Push'' method
investigated in this paper is that it tries to avoid copying
data values (as far as possible).
In the standard implementation of bottom-up evaluation,
the result of applying a rule
is stored in a derived relation.
In contrast,
Prolog implementations do not copy variable values
or materialize derived literals
(except with tabling).
It might be that the ``Push'' method
makes bottom-up evaluation competitive because it gives it something	
which Prolog systems already had.
However,
as we will see,
good performance depends on many factors
(including, e.g., efficient data structures).
The message of this paper is certainly not
that the ``Push'' method alone could make us win benchmarks.





\section{Query Language, General Setting}

\label{secQueryLanguage}

In this paper,
we consider the very basic case of Datalog,
i.e.~pure Prolog without negation
and without function symbols
(i.e.~terms can only be variables or constants).
A logic program is a set of rules,
for example
\begin{center}
\begin{tt}
\begin{tabular}{@{}l@{}}
answer(X) :- grandparent(sam, X).\\
grandparent(X, Z) :- parent(X, Y), parent(Y, Z).
\end{tabular}
\end{tt}
\end{center}
This computes the grandparents of~\code{sam},
given a database relation~\code{parent}.
For our bottom-up method,
we assume that there is a ``main'' predicate \code{answer},
for which we want to compute all derivable instances.
For the other systems,
the body of the \code{answer}-rule is directly posed as query/goal.

We require range-restriction (allowedness),
i.e.~all variables in the head of the rule
must also appear in a body literal.
Therefore,
when rules are applied from right (body) to left (head),
only facts are derived,
i.e.~variable-free atomic formulas
like \code{grandparent(sam,~john)}.

As usual in deductive databases,
the predicates are classified into
\begin{itemize}
\item
EDB-predicates (``extensional database''),
which are defined only by facts
(usually a large set of facts
stored in a database,
or maybe specially formatted files).
In the above example,
\code{parent} is an EDB predicate.
\item
IDB-predicates (``intensional database''),
\kern-0.8pt
which are defined by rules.
\kern-0.8pt
In the 
example,
\code{grandparent} and \code{answer}
are IDB predicates.
\end{itemize}

The execution of a logic program/query has three distinct phases:
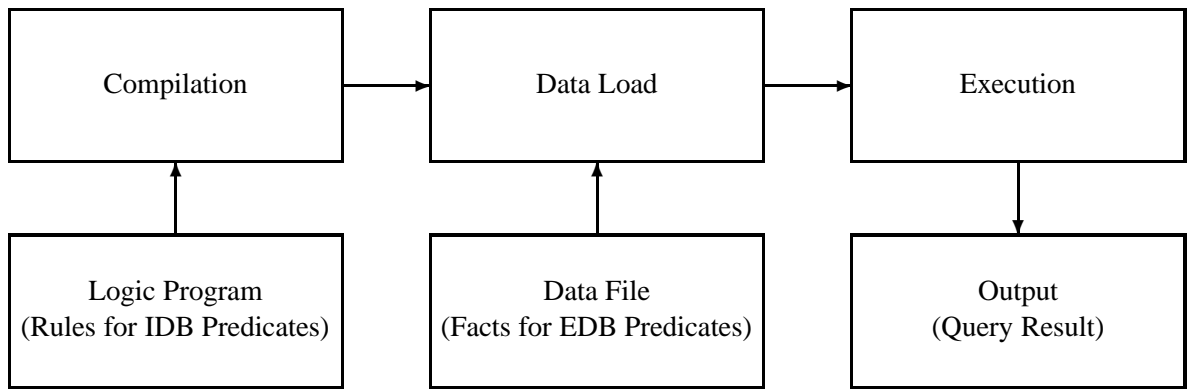
\begin{center}
\setlength{\unitlength}{1mm}
\thicklines
\begin{picture}(156,50)(0,0)
\put(  0, 0){\framebox(44,20){\begin{tabular}{@{}c@{}}
				Logic Program\\
				(Rules for IDB Predicates)
				\end{tabular}}}
\put( 22,20){\vector(0,1){10}}
\put(  0,30){\framebox(44,20){\begin{tabular}{@{}c@{}}
				Compilation
				\end{tabular}}}
\put( 44,40){\vector(1,0){12}}
\put( 56, 0){\framebox(44,20){\begin{tabular}{@{}c@{}}
				Data File\\
				(Facts for EDB Predicates)
				\end{tabular}}}
\put( 78,20){\vector(0,1){10}}
\put( 56,30){\framebox(44,20){\begin{tabular}{@{}c@{}}
				Data Load
				\end{tabular}}}
\put(100,40){\vector(1,0){12}}
\put(112, 0){\framebox(44,20){\begin{tabular}{@{}c@{}}
				Output\\
				(Query Result)
				\end{tabular}}}
\put(134,30){\vector(0,-1){10}}
\put(112,30){\framebox(44,20){\begin{tabular}{@{}c@{}}
				Execution
				\end{tabular}}}
\end{picture}
\end{center}
The input to our transformation are the rules for the IDB-predicates,
i.e.~these are known at ``compile time''.
The concrete data for the EDB-predicates
are only known at ``run time''.
Therefore,
the \CPP\ program resulting from our translation
can be executed for different database states
(sets of facts).
In this way,
optimizations or precomputations done at compile time
can be amortized over many program executions.
Actually,
since the database is usually large
and the program relatively small,
even a single execution might suffice
so that work invested during the transformation ``pays off''.


In the OpenRuleBench tests,
the runtime is again split into the time required
for loading the data file
and the real execution time
when the data is already in main memory
(and useful data structures have been built).
In the OpenRuleBench paper~\cite{LFWK09},
the authors are interested only in the time for the inference
(i.e.~the execution phase),
and only this time is shown in the paper.
We show the times for both,
the data load and the execution,
and the overall runtime measured externally.
Only in case of DLV	
it was not possible to separate this
(we did measure the time for only loading the data,
but it is not clear how much of the data structure building
is done if the data is not used).

Of course,
handling the data load separately
is interesting if many queries are executed later
on the data in main memory,
i.e.~if one had a main memory DBMS server.

For the most part,
we did not consider the compilation time
which might be considered as giving us an unfair advantage.
Systems which have no separate compilation phase
cannot profit from this.



\section{Data Structures}

\label{secDataStructures}

When the facts from a data file are loaded,
they are stored in main-memory relations.
A large part of the \CPP\ library
which is used in the translation result
implements various data structures for main-memory relations.
Most relations are \CPP\ templates
which are parameterized with the tuple type.
However,
there are a few exceptions
(experimental/old implementations, or data structures for very special cases).

Actually,
all arguments are currently mapped to integers.
We have implemented a string table
using a radix tree similar to~\cite{LKN13}
in order to assign unique, sequential numbers to strings.
If we know that different arguments have different domains
(which are never joined)
we use different such ``string tables''
to keep the numbers dense
(which is good for array based data structures).

There are basically three types of relation data structures:
\begin{itemize}
\item
Lists,
which support access to the tuples
with binding pattern ``\code{ff\ldots f}'',
i.e.~all arguments are free (output arguments).
One can open a cursor over the list and iterate over all its elements (tuples).
Lists are implemented as a tree of 4~KB pages,
each containing an array of tuples
(i.e.~the tuples are stored in consecutive memory locations
which helps to improve CPU cache utilization).
The next level of the tree contains pages with pointers
to the data pages
(similar to the management of blocks in the original Unix file system).
\item
Sets,
which support access to the tuples
with binding pattern ``\code{bb\ldots b}'',
i.e.~all arguments are bound (input arguments).
One can insert a tuple and gets the information
whether the insertion was successful (new tuple)
or not (the tuple was contained in the set already).
We are experimenting with different set implementations,
currently:
(a)~a simple hash table of fixed size,
(b)~a dynamic hash table which doubles its size if it gets too full,
and (c)~an array of bitmaps
(some of the benchmarks have only a small domain of integers from~0 to~1000).
\item
Maps (or really multimaps) for all other cases,
e.g.~a binding pattern like~\code{bf}:
Given values for the input arguments
(in this case, the first one),
one can open a cursor over the corresponding (tuples of) values
for the output arguments.
The current state is that we have only a flexible array implementation
for a single integer input argument,
but a general map will be implemented soon.
\end{itemize}

Temporary relations might be needed during execution
of a logic program:
If a rule has more than one IDB body literal,
previously derived facts matching these literals must be stored
(except possibly one of the literals,
if we know that its facts arrive last
and we can immediately process them).
In this case it is important
that when a cursor is opened,
the set of tuples it loops over is not affected by future insertions.

Temporary sets are also needed for duplicate elimination
during query evaluation.
For some benchmarks,
this has a big influence on speed.



\section{The Push Method for Bottom-Up Evaluation}

\label{secPushMethod}

The push method has been introduced in~\cite{Bra10CPP},
and a version with partial evaluation
has been defined in~\cite{BS15WLP}.
The following is just a quick reminder
in order to be able to understand different variants of the code.

The basic method only works with rules with at most one IDB body literal
(as generated, e.g., by the SLDMagic method~\cite{Bra00}).
Rules with multiple IDB body literals
are normalized by creating intermediate storage
for previously derived tuples
(similar to seminaive evaluation with a single tuple as ``delta'' relation,
see~\cite{BS15WLP}).

In the push method,
a fact is represented
by a position in the program code
and values of \CPP\ variables.
A fact type is a positive IDB literal
with constants and \CPP\ variables as arguments,
for instance \code{p(a,~v{\U}1,~v{\U}2)}.
If execution reaches a code position corresponding to this fact type,
the corresponding fact with the current values
of the variables \code{v{\U}1} and \code{v{\U}2}
has been derived.
Now this derived fact is immediately ``pushed''
to rules with a matching body literal.
A rule application consists of
\begin{itemize}
\item
a rule of the given program,
\item
a fact type that is unifiable with the IDB body literal of this rule,
or a special marker ``\code{init}'' if the rule has only EDB body literals,
\item
a fact type that is general enough to represent
all facts which can be derived with the rule
applied to the input fact type.
\end{itemize}
A rule application graph consists of the fact types as nodes
(plus an ``\code{init}'' node),
and the rule applications as edges.

One can use different sets of fact types.
In~\cite{Bra10CPP} the fact types had the form
\code{p(p{\U}1,~p{\U}2,~\ldots)},
i.e.~there was one \CPP\ variable
for each argument position of each IDB predicate.
This requires a lot of copying in a rule application.
The approach of~\cite{BS15WLP} did avoid this by computing
a set of fact types as a fixpoint of an abstract~\m{\TP}-operator
working with \CPP\ variables instead of real constants.
These variables were introduced if a data value was needed
from an EDB body literal (the real values are not known at compile time).
This avoids copying,
but the set of generated fact types can ``explode''
(as, e.g., in the wine ontology benchmark).
Therefore,
our current implementation supports both approaches,
and we work on mixed variants which combine the advantages of both.

There is one code piece for each rule application.
At the end of this code piece,
a fact has been derived and the variables of the resulting fact type
have been set.

Of course,
a derived fact might match body literals of several rules.
In that case control jumps to one such rule application,
and the other possibilities are later visited
via backtracking.
If variables are changed when there might still be backtrack points
on the stack which need the old value,
these are saved on a stack, too.
In both of the above variants of the push method,
this happens only for recursive rule applications.

It is also possible that a rule application can produce several facts.
Only one is computed at a time and immediately used,
other facts are computed later via backtracking.
Therefore,
the code piece for a rule application has two entry points:
One for computing the first fact (``\code{START}''),
and one for computing an additional fact (``\code{CONT}'').
Before control jumps to a place where the derived fact is used,
the ``\code{CONT}''-label is put on the backtrack stack.
In case a rule application is unsuccessful
or there is no further fact that can be derived, 
the main loop is continued, which takes the next rule application
from the backtrack stack and executes its code piece.

\section{About the Time and Memory Measurements}

\label{secBenchmark}

We compared our method with the following systems:
\begin{itemize}
\item
XSB Prolog version 3.6 (Gazpatcho)
\item
YAP Prolog version 6.2.2
\item
DLV version x86-64-linux-elf-static
\end{itemize}

Both Prolog systems support tabling.
The DLV system uses a variant of magic sets transformation
and then does bottom-up evaluation.

We executed the benchmarks on
a machine with
two 6272 AMD Opteron 
16~Core~CPUs.
However,
the current version of our program does not use multiple threads
(we will consider this in future research).
The machine has 64~GB of RAM.
This is quite a lot,
but we also measure how much memory is really used by the systems.
The operating system is Debian \code{x86{\U}64} GNU/Linux~3.2.63.

The overall execution time (``elapsed wall clock time'')
and the memory (``maximum resident set size'')
for each test
was measured with the Linux \code{time} program.
The time for loading the data and for executing the query
are measured by functions of each system.

Every test was run ten times and the time average values were calculated. 
Additionally, for the XSB and YAP systems the times for loading the
data and program files and for executing the queries were separately 
measured. 

For the DLV system the use of the \code{-stats} option is officially
discouraged for benchmark purposes, so for estimating the loading time
the loading of the data files was measured once using the 
\code{time}
program; the value appears in parentheses in the tables.

XSB Prolog compiles the input files to a \code{xwam} file which can
be loaded quite fast, but for better comparison with the other systems dynamic
loading was used.



\section{The DBLP Benchmark}

\label{secDBLP}

For the DBLP benchmark~\cite{LFWK09},
data from the DBLP bibliography
(\code{dblp.uni-trier.de})
is stored as a large set of facts
for an EAV-relation
\begin{center}
\code{att(DocID, Attribute, Value)}.
\end{center}
The file contains slightly more than 2.4~million facts of this type,
for instance:
\begin{center}
\code{att('node12eo5mnsvx56023',title,'On the Power of Magic.').}
\end{center}
It is about 122~MB large.
The test query is
\begin{center}
\begin{tt}
\begin{tabular}{@{}l@{}}
answer(Id, T, A, Y, M) :-\\
~~~~att(Id, title,~ T),\\
~~~~att(Id, year,~~ Y),\\
~~~~att(Id, author, A),\\
~~~~att(Id, month,~ M).
\end{tabular}
\end{tt}
\end{center}
In this case,
\code{att} is an EDB-predicate,
and \code{answer} is the only IDB-predicate.
The file contains data of about 215.000~documents
(i.e.~11.3~facts per document).%
\footnote{%
The DBLP data are available as XML file from
\url{http://dblp.uni-trier.de/xml/}.
The DBLP has now data of 3.4~million documents,
the XML-file is 1.8~GB large.
We plan to transform this larger data file
to the same type of Datalog facts
as the official OpenRuleBench DBLP test file.}

It seems that the key to success for this benchmark problem
are the data structures to represent the facts.
When the data is loaded,
we must store it in relation data structures
for later evaluation of the EDB body literals.
Note that the program/query is known
when the data is loaded,
so we can try to create an optimal index structure for each EDB body literal.
In particular,
selections for constants known at compile time
can already be done when the data is loaded.
In the example,
all body literals contain different constants,
therefore we create a distinct relation
for each of them.

E.g.~there is a relation~\code{att{\U}title},
which represents \code{att}-facts
that match 
the first body literal
\code{att(Id,~title,~T)}.
Of course, the constant is not explicitly stored,
therefore the relation has only two columns.
We use a nested loop/index join.
The first occurrence of each variable binds that variable,
and at all later occurrences,
its value is known.
Since this is the first body literal,
both variables are still free,
and the relation \code{att{\U}title}
is accessed with binding pattern~\code{ff},
i.e.~it is a list.

For the second body literal \code{att(Id,~year,~Y)},
there is a relation~\code{att{\U}year}
which represents only \code{att}-facts with \code{year} as second argument.
Again,
this argument is not explicitly represented in the stored tuples.
Since the value of the variable~\code{Id} is known
when this body literal is evaluated,
the relation must support the binding pattern~\code{bf}.
Therefore it is a map
(or really multimap, since we do not know that there is only one year
for each document).

In the same way,
the other two body literals are turned into exactly matching relations
of type~\code{bf}.
This means that although the loader has to look at all facts in the data file,
it stores only those facts which are relevant for the query
(which are 
35\% of the facts in the data file).

The strings that occur in the data are mapped to unique, sequential integers.
Since the three arguments of the predicate \code{att} are never compared,
we use three distinct ``string tables'':
This makes it simpler to implement the maps based on a kind of flexible array
--- for other map data structures, this would probably have no advantage.
However,
also bitmap implementations of sets
profit from a small, dense domain.
In summary,
the main data structures are:
\vspace*{1mm}
\begin{center}
\begin{tabular}{@{}|l|r|r|@{}}
\hline
Data Structure&Rows/Strings&Memory (KB)\\
\hline
\hline
String table for Arg.~1&214\kern1pt 905&91\kern1pt 612\\
String table for Arg.~2&19&4\\
String table for Arg.~3&923\kern1pt 881&284\kern1pt 052\\
\hline
\hline
List for \code{att(Id,~title,~T)}&209\kern1pt 943&1\kern1pt 648\\
Map for \code{att(Id,~year,~Y)}&209\kern1pt 944&844\\
Map for \code{att(Id,~author,~A)}&440\kern1pt 934&3\kern1pt 688\\
Map for \code{att(Id,~month,~A)}&2\kern1pt 474&844\\
\hline
\end{tabular}
\end{center}
\vspace*{1mm}



One might expect that creating these indexes slows down
the loading of data
(in effect, some time is moved from the execution phase to the load phase).
However,
our loader is quick.
Here are the benchmark results:
\vspace*{1mm}
\begin{center}
\begin{tabular}{|l|r|r|r|r|r|r|}
\hline
System&Load (ms)&Execution (ms)&Total time (ms)&Factor&Memory (KB)&Mem.~Diff.\\
\hline
\hline
Push&2\kern1pt 565\phantom{)}&22&2\kern1pt 610&1.0
		&385\kern1pt 172
		&383\kern1pt 990\\
XSB&89\kern1pt 045\phantom{)}&2\kern1pt 690&92\kern1pt 390&35.4
		&415\kern1pt 500
		&404\kern1pt 535\\
XSB (trie)
&90\kern1pt 710\phantom{)}&269&91\kern1pt 275&35.0
		&380\kern1pt 259
		&369\kern1pt 294\\
	YAP&24\kern1pt 878\phantom{)}&7\kern1pt 370&32\kern1pt 438&12.4
		&813\kern1pt 760
		&808\kern1pt 911\\
	DLV&(20\kern1pt 110)&---&25\kern1pt 898&9.9
		&926\kern1pt 864
		&926\kern1pt 167\\
\hline
\end{tabular}
\end{center}
\vspace*{1mm}
The total time is dominated by the loading time,
however the pure execution time is also interesting.
By the way,
just reading the file character by character with the standard library
takes about the same time as our loader
(which reads the file in larger chunks).
In the OpenRuleBench programs,
XSB did not use the trie index,										
but this dramatically improves execution time						
(the influence on the total time is not big).

The main memory consumption of our ``Push'' implementation is acceptable
(383~MB, which is about the same as XSB,
and less than half of YAP and XSB).
However,
since only one third of the facts are actually stored,
and only with a specific binding pattern,
memory could become an issue for other application scenarios.

In order to check how much memory was used by the data for the benchmark
and how much is program code in the system
(including possibly large libraries)
we determined the memory used by just starting and stopping
each system (without loading data or compiling rules).
The result is:
\begin{center}
\begin{tabular}{@{}|l|r|@{}}
\hline
System&Base Memory (KB)\\
\hline\hline																	
Push&1\kern1pt 182\\
XSB&10\kern1pt 965\\
YAP&4\kern1pt 849\\
DLV&697\\
\hline
\end{tabular}
\end{center}
The column ``Mem.~Diff.'' in the benchmark result table
contains the difference of the memory used in the benchmark
minus this baseline.




Since this query is a standard database task,
we also tried HSQLDB~2.3.4.
It loads all data into main memory when the system starts,
this took 32\kern1pt 471~ms.
Executing the following SQL query took~3\kern1pt 863~ms:
\begin{center}
\begin{tt}
\begin{tabular}{@{}l@{ }l@{}}
SELECT&COUNT(*)\\
FROM  &ATT ATT{\U}TITLE,
      ATT ATT{\U}YEAR,
      ATT ATT{\U}AUTHOR,
      ATT ATT{\U}MONTH\\
WHERE &ATT{\U}TITLE.ATTRIBUTE = 'title'\\
AND   &ATT{\U}YEAR.ATTRIBUTE  = 'year'\\
AND   &ATT{\U}AUTHOR.ATTRIBUTE  = 'author'\\
AND   &ATT{\U}MONTH.ATTRIBUTE  = 'month'\\
AND   &ATT{\U}TITLE.DOC{\U}ID = ATT{\U}YEAR.DOC{\U}ID\\
AND   &ATT{\U}TITLE.DOC{\U}ID = ATT{\U}AUTHOR.DOC{\U}ID\\
AND   &ATT{\U}TITLE.DOC{\U}ID = ATT{\U}MONTH.DOC{\U}ID
\end{tabular}
\end{tt}
\end{center}

The total time for starting the database, executing the query
and shutting down the database was 40\kern1pt 777~ms
(the shutdown took 4\kern1pt 330~ms).
HSQLDB did use three threads (300\%~CPU)
and 3~GB of main memory
(i.e.~three times the system resources of the deductive systems above).
HSQLDB is written in Java
(and was executed in the OpenJDK 64-bit Server VM: IcedTea~2.6.6, Java~1.7.0).
We used an index over \code{(ATTRIBUTE, DOC{\U}ID)}.
We tried also an index with the two attributes inversed,
or all three attributes,
but all this did not change much.
Although there might be further options for tuning HSQLDB,
it is at least not easy
to get much better runtimes with a relational database.



\section{The Join~1 Benchmark}

\label{secJoinOne}

The Join1 example from~\cite{LFWK09}
contains the following rules:
\begin{center}
\begin{tt}
\begin{tabular}{@{}l@{}}
	a(X, Y)~ :- b1(X, Z), b2(Z, Y).\\
	b1(X, Y) :- c1(X, Z), c2(Z, Y).\\
	b2(X, Y) :- c3(X, Z), c4(Z, Y).\\
	c1(X, Y) :- d1(X, Z), d2(Z, Y).
\end{tabular}
\end{tt}
\end{center}
The EDB predicates are \code{c2}, \code{c3}, \code{c4}, \code{d1}, \code{d2}.
There are three data files:
One with 10\kern0.2em 000~facts each
(i.e.~50\kern0.2em 000~facts in total),
one with 50\kern0.2em 000~facts each,
and one with 250\kern0.2em 000~facts each.
The data values are randomly generated integers between~1 and~1000.
Different queries are considered in~\cite{LFWK09},
namely \code{a(X,~Y)}, \code{b1(X,~Y)}, \code{b2(X,~Y)},
and the same with bound first or second argument,
e.g.~the query~\code{a(1,~Y)}.
In our test,
we only tried the query~\code{a(X,~Y)}.

The benchmark results for the small data file
(10\kern0.2em 000~facts per EDB-predicate, 624~KB) are:
\vspace{-1mm}
\begin{center}
\begin{tabular}[t]{|l|r|r|r|r|r|}
\hline
System&Load (ms)&Execution (ms)&Total time (ms)&Factor&Memory (KB)\\
\hline
\hline
Push (Bitmap)&14\phantom{)}&1\kern0.2em 772&1\kern0.2em 787&1.0
		&7\kern0.2em 311\\
Push (Dyn.Hashtab)&11\phantom{)}&10\kern0.2em 372&10\kern0.2em 383&5.8
		&45\kern0.2em 688\\
xsb-btc&141\phantom{)}&22\kern0.2em 432&22\kern0.2em 827&12.8
		&82\kern0.2em 667\\
XSB		&148\phantom{)}&24\kern0.2em 551&25\kern0.2em 002&14.0
		&82\kern0.2em 526\\
YAP&421\phantom{)}&17\kern0.2em 557&18\kern0.2em 067&10.1
		&10\kern0.2em 933\\
DLV&&&147\kern0.2em 172&82.4
		&461\kern0.2em 646\\
\hline
\end{tabular}
\end{center}
\vspace{2mm}
The benchmark scripts (from the authors of~\cite{LFWK09})
used a version of XSB with ``batched scheduling''
for this test.
However,
it gives only a relatively small improvement over the standard version
(with ``local scheduling'').

The domain of values are in all cases integers from~1 to~1000,
so many duplicates will be generated.
Actually,
this test is dominated by the time for duplicate elimination.
With an ``array of bitmaps'' implementation,
we are much faster than with a dynamic hash table
(it doubles its size if the chains get too long).
Here are the relations used in our Push implementation:
\vspace{2mm}
\begin{center}
\begin{tabular}{@{}|l|l|r|r|@{}}
\hline
Table&Comment&Rows&Memory (KB)\\
\hline
\hline
\code{c2{\U}bf}&EDB-Relation (Map)&10\kern0.2em 000&88\\
\code{c3{\U}ff}&EDB-Relation (List)&10\kern0.2em 000&84\\
\code{c4{\U}bf}&EDB-Relation (Map)&10\kern0.2em 000&88\\
\code{d1{\U}ff}&EDB-Relation (List)&10\kern0.2em 000&84\\
\code{d2{\U}bf}&EDB-Relation (Map)&10\kern0.2em 000&88\\
\hline
\code{b1{\U}fb}&Temporary IDB-Relation (Map)
	&634\kern0.2em 088
	&4\kern0.2em 976\\
\hline
\code{b1{\U}bb}&Duplicate Check for IDB-Pred.~(Set)
	&634\kern0.2em 088
	&14\kern0.2em 004\\
\code{b2{\U}bb}&Duplicate Check for IDB-Pred.~(Set)
	&95\kern0.2em 954
	&2\kern0.2em 524\\
\code{c1{\U}bb}&Duplicate Check for IDB-Pred.~(Set)
	&95\kern0.2em 570
	&2\kern0.2em 520\\
\code{a{\U}bb}&Duplicate Check for IDB-Pred.~(Set)
	&1\kern0.2em 000\kern0.2em 000
	&19\kern0.2em 724\\
\hline
\end{tabular}
\end{center}
\vspace{2mm}
In the version with bitmap duplicate check,
the last four sets need only 128~KB each.
The temporary IDB-relation is needed because
the first rule has two IDB-body literals:
\begin{center}
a(X, Y) :- b1(X, Z), b2(Z, Y).\\
\end{center}
Therefore,
we compute first the \code{b1}-facts and store them in~\code{b1{\U}fb}.
Later,
when we derive a \code{b2}-fact,
we use it immediately to derive \code{a}-facts with this rule.
Since in this case \code{Z} is bound from the \code{b2}-fact,
we need the binding pattern~\code{fb}
for the intermediate storage of~\code{b1}-facts.

If one does not do any duplicate elimination,
one gets 99.9~million result tuples,
i.e.~on average every answer is computed 100~times.
Our implementation of this needed 2\kern0.2em 501~ms execution time,
i.e.~was quite fast.
This shows again that this benchmark depends a lot
on the efficiency of duplicate elimination.

Our results with the hash table are not impressive and could be fully explained
with using native code instead of emulating an abstract machine.
However,
it scales better than XSB as the other data sets show.
XSB timeouts already on the middle data set with 50\kern0.2em 000~rows
per EDB predicate
(timeout is set at 30~minutes).
Our Push implementation (with hash table) needs 106~s.
This means that it is better by at least the factor~17.

However, we found that tabling all IDB predicates for XSB and YAP and introducing
indexes for the EDB relations in XSB dramatically improved the 
runtime performance of
these systems (in the original program files, tabling was only done 
for the predicate \code{a/2} in XSB, and no indices were used). This led to
the following results:
\vspace{-1mm}
\begin{center}
\begin{tabular}[t]{|l|r|r|r|r|r|}
\hline
System&Load (ms)&Execution (ms)&Total time (ms)&Memory (KB)\\
\hline
\hline
XSB&414\phantom{)}&11\kern0.2em 302&11\kern0.2em 927
		&123\kern0.2em 885\\
YAP&437\phantom{)}&6\kern0.2em 719&7\kern0.2em 256
		&133\kern0.2em 622\\
\hline
\end{tabular}
\end{center}
\vspace{2mm}
Due to creating indexes, the loading time of XSB has increased, and both 
systems need more memory for the tables. 
\section{The Transitive Closure Benchmark with Both Arguments Free}

\label{secTCFF}

An important test for a deductive database system is the transitive closure
program:
\begin{center}
\begin{tt}
\begin{tabular}{@{}l@{}}
	tc(X, Y) :- par(X, Y).\\
	tc(X, Y) :- par(X, Z), tc(Z, Y).
\end{tabular}
\end{tt}
\end{center}
Of course,
this is one of the OpenRuleBench benchmark problems for recursion.
It uses 10~data files of different size (see below),
we first concentrated on the file
\code{tc{\U}d1000{\U}parsize50000{\U}xsb{\U}cyc.P}
where
the arguments of the \code{par}-relation are taken
from a domain of 1000~integers (1~to~1000)
(i.\,e.\ the graph has 1000~nodes);
there are 50\kern0.2em 000~randomly generated \code{par}-facts
(i.\,e.\ the graph has 50\kern0.2em 000~edges);
the graph is cyclic;
%
and the file size is 673~KB.
%
%
OpenRuleBench used three test queries:
\code{tc(X,\kern0.2em Y)},
\kern0.3em
\code{tc(1,\kern0.2em Y)},
\kern0.3em
and \code{tc(X,\kern0.2em 1)}%
\footnote{Actually, anonymous variables were used in the test,
because the answers are not further processed.}.
Here are the benchmark results
for the first query asking for the entire \code{tc} relation
(i.\,e.\ with binding pattern~\code{ff}):					
\vspace{1mm}
\begin{center}
\begin{tabular}{@{}|l|r|r|r|r|r|@{}}
\hline
System&Load (ms)&Execution (ms)&Total time (ms)&Factor&Memory (KB)\\
\hline
\hline
Push&11\phantom{)}&2\kern0.2em 171&2\kern0.2em 190&1.0
		&21\kern0.2em 832\\
Seminaive&10\phantom{)}&4\kern0.2em 598&4\kern0.2em 613&2.1
		&29\kern0.2em 587\\
XSB&1\kern0.2em 095\phantom{)}&11\kern0.2em 357&12\kern0.2em 641&5.8
		&134\kern0.2em 024\\
YAP&424\phantom{)}&19\kern0.2em 407&19\kern0.2em 867&9.1
		&145\kern0.2em 566\\
DLV&(260)&---&109\kern0.2em 673&50.1
		&404\kern0.2em 737\\
\hline
\end{tabular}
\end{center}
\vspace{1mm}
Obviously,
one must detect duplicates in order to ensure termination.
But even with non-cyclic data,
there are many duplicates:
Each node is linked to~5\% of the other nodes.
So the probability is quite high
that even after short paths the same node is reached multiple times.	
XSB uses tabling 
to solve this problem~\cite{SSW94,CW96SLG,Swi99}.
The relations we used for this problem are:
\begin{center}
\begin{tabular}{@{}|l|l|r|r|@{}}
\hline
Table&Comment&Rows&Memory (KB)\\
\hline
\hline
\code{par{\U}ff}&EDB-relation (as list), used in first rule&50000&396\\
\code{par{\U}fb}&EDB-relation (as map), used in second rule&50000&400\\
\code{tc{\U}bb}&Duplicate check for result (set)&1000000&19\kern0.2em 536\\
\hline
\end{tabular}
\end{center}
The first rule produces initial \code{tc}-facts
and simply loops over the entire \code{par}-relation,				
therefore it needs a version implemented as list.
The second (recursive) rule is activated in the ``Push'' method
when a new \code{tc}-fact is derived. 
Therefore,
a value for variable~\code{Z} is known
and the EDB-relation is accessed with binding pattern~\code{fb}.

We also implemented a standard semina\"ive bottom-up evaluation
for comparison.
It used the same data structures as our Push implementation,
it only needed an additional list \code{tc{\U}ff} for the result.
It is slower than the ``Push'' method.
Probably, this is due to the different memory accesses:
In each iteration, freshly derived tuples are first stored in the result list,
and they are accessed again in the next iteration.

The following table
lists the benchmark results for the full set of 10~data files
of various size and characteristics.
We compared only the execution times with XSB.
We used here the original settings for XSB from the OpenRuleBench scripts.
In the result shown above,
a \code{trie}-index was added,
which improved the performance (12.6\kern0.2em s instead of 15.5\kern0.2em s).
\begin{center}
\begin{tabular}{@{}|r|r|c|r|r|r|@{}}
\hline
\multicolumn{1}{@{}|c}{\vrule height 1.05em depth 0.5em width 0pt Rows}&
	\multicolumn{1}{|c}{Dom}&
	\multicolumn{1}{|c}{cyc}&
	\multicolumn{1}{|c}{XSB}&
	\multicolumn{1}{|c}{Push}&
	\multicolumn{1}{|c|@{}}{Factor}\\ 
\hline
\hline
50\kern0.2em 000&1000&N&
	5.980\kern0.1em s
	&0.732\kern0.1em s
	&8.2
	\\
50\kern0.2em 000&1000&Y&
	15.513\kern0.1em s
	&2.160\kern0.1em s
	&7.2
	\\
250\kern0.2em 000&1000&N&
	33.274\kern0.1em s
	&4.350\kern0.1em s
	&7.6
	\\
250\kern0.2em 000&1000&Y&
	82.497\kern0.1em s
	&12.440\kern0.1em s
	&6.6
	\\
500\kern0.2em 000&1000&N&
	76.673\kern0.1em s
	&10.230\kern0.1em s
	&7.5
	\\
500\kern0.2em 000&1000&Y&
	187.200\kern0.1em s
	&30.620\kern0.1em s
	&6.1
	\\
500\kern0.2em 000&2000&N&
	139.881\kern0.1em s
	&23.870\kern0.1em s
	&5.9
	\\
500\kern0.2em 000&2000&Y&
	329.873\kern0.1em s
	&61.460\kern0.1em s
	&5.4
	%
	%
	\\
\kern-3pt 1\kern0.2em 000\kern0.2em 000&2000&N&
	297.870\kern0.1em s
	&56.623\kern0.1em s
	&5.3
	\\
\kern-3pt 1\kern0.2em 000\kern0.2em 000&2000&Y&
	714.721\kern0.1em s
	&150.270\kern0.1em s
	&4.8
	%
	\\
\hline
\end{tabular}
\end{center}
\vspace{2mm}
The first column gives the number of rows in the data\-base,
i.\,e.\ the size of the \code{par}-relation.
The second column lists the size of the domain of values
which occur in the \code{par} columns.%
\footnote{Some of the graphs which claim to be non-cyclic
actually do contain cycles.
The largest non-cyclic graph
with 1000~nodes is \m{\setCond{(i,j)}{1\le i\lt j\le 1000}}.
This has only \m{999*1000/2=499\:500}~edges.}
The last column shows the speed improvement factor
of our Push implementation over XSB.
For loading the last relation (1~million rows, 14.2~MB),
XSB needs 90.2\kern0.1em s,
our Push implementation needs 0.2\kern0.1em s.




\section{The Transitive Closure Benchmark with the First Argument Bound}

\label{secTCBF}
In this benchmark,
one is not interested in all connected pairs in the transitive closure,
but only in nodes reachable from node~\code{1}.
So the binding pattern for accessing the predicate is~\code{bf}.

Different systems use different methods to pass bindings
to called predicates.
E.g.~for XSB and YAP,
this is a feature of SLD-resolution.
For a system based on bottom-up evaluation,
the magic set method is the standard solution~\cite{BMSU86,BR87}.
Probably DLV does this.
However, since magic sets are known
to have problems with tail recursions~\cite{Ros91},
we use SLDMagic~\cite{Bra00} for our ``Push'' method instead
(the ``Push'' method alone is a pure bottom-up method
would not pass query bindings).
The output of the transformation is:
\begin{center}
\begin{tt}
\begin{tabular}{@{}l@{}}
p1(A) :- p1(B), par(B,A).\\
p1(A) :- par(1,A).\\
p0(A) :- p1(B), par(B,A).\\
p0(A) :- par(1,A).\\
tc(1,A) :- p0(A).
\end{tabular}
\end{tt}
\end{center}
The predicate \code{p0} is not really needed,
but since the SLDMagic prototype produces this program,
it would be unfair to improve it manually.
Nevertheless,
it can be evaluated extremely fast,
since it reduced the arity of the recursive predicate.
Here are the benchmark results:
\vspace*{1mm}
\begin{center}
\begin{tabular}{@{}|l|r|r|r|r|r|@{}}
\hline
System&Load (ms)&Execution (ms)&Total time (ms)&Factor&Memory (KB)\\
\hline
\hline
Push&10\phantom{)}&14&30&1.0
	&9\kern0.2em 716\\
XSB&1\kern0.2em 098\phantom{)}&7\kern0.2em 142&8\kern0.2em 418&280.6
	&86\kern0.2em 936\\
YAP&440\phantom{)}&8\kern0.2em 743&9\kern0.2em 246&308.2
	&91\kern0.2em 325\\
DLV&(260)&---&110\kern0.2em 779&3\kern0.2em 692.6
	&404\kern0.2em 736\\
\hline
\end{tabular}
\end{center}
\vspace*{1mm}
At least the standard version of tabling
has the same problem with tail recursions
as magic sets:
One cannot materialize the derived \code{tc}-facts in this case,
since they are not only for the type \code{tc(1,\kern0.2em Y)},
but all facts \code{tc(X,\kern0.2em Y)}
for every~\code{X} reachable from~\code{1}.
Depending on the graph,
this can make the difference between a linear number of facts
and a quadratic number of facts.
XSB and YAP might be affected by this problem.

It could 
be argued that the 
good results for the Push Method are mainly due to the SLDMagic program
transformation.
Therefore,
in a further test,
the SLDMagic-transformed program was
also given as input to the other systems.
Indeed, they did profit 
from the transformation
(in runtime as well as in memory usage), 
but the Push Method is still 
in the first place:
\vspace*{1mm}
\begin{center}
\begin{tabular}{@{}|l|r|r|r|r|r|@{}}
\hline
System&Load (ms)&Execution (ms)&Total time (ms)&Factor&Memory (KB)\\
\hline
\hline
Push&10\phantom{)}&14&30&1.0
	&9\kern0.2em 716\\
XSB&1\kern0.2em 093\phantom{)}&10&1\kern0.2em 292&43.1
	&14\kern0.2em 298\\
YAP&449\phantom{)}&36&562&18.7
	&16\kern0.2em 696\\
DLV&(260)&---&389&13.0
	&11\kern0.2em 997\\
\hline
\end{tabular}
\end{center}
\vspace*{1mm}
For the other data files,
we compared the execution times only with XSB
(using SLDMagic for the Push method, but not for XSB):
\begin{center}
\begin{tabular}{@{}|r|r|c|r|r|r|@{}}
\hline
\multicolumn{1}{@{}|c}{\vrule height 1.05em depth 0.5em width 0pt Rows}&
	\multicolumn{1}{|c}{Dom}&
	\multicolumn{1}{|c}{cyc}&
	\multicolumn{1}{|c}{XSB}&
	\multicolumn{1}{|c}{Push}&
	\multicolumn{1}{|c|@{}}{Factor}\\ 
\hline
\hline
50\kern0.2em 000&1000&N
	&1.296\kern0.1em s
	&0.010\kern0.1em s
	&130
	\\
50\kern0.2em 000&1000&Y
	&6.912\kern0.1em s
	&0.010\kern0.1em s
	&691
	\\
250\kern0.2em 000&1000&N
	&9.309\kern0.1em s
	&0.030\kern0.1em s
	&310
	\\
250\kern0.2em 000&1000&Y
	&35.098\kern0.1em s
	&0.030\kern0.1em s
	&1170
	\\
500\kern0.2em 000&1000&N
	&19.989\kern0.1em s
	&0.050\kern0.1em s
	&400
	\\
500\kern0.2em 000&1000&Y
	&69.929\kern0.1em s
	&0.060\kern0.1em s
	&1165
	\\
500\kern0.2em 000&2000&N
	&36.067\kern0.1em s
	&0.060\kern0.1em s
	&601
	\\
500\kern0.2em 000&2000&Y
	&154.110\kern0.1em s
	&0.070\kern0.1em s
	&2202
	\\
\kern-3pt 1\kern0.2em 000\kern0.2em 000&2000&N
	&80.117\kern0.1em s
	&0.130\kern0.1em s
	&616
	\\
\kern-3pt 1\kern0.2em 000\kern0.2em 000&2000&Y
	&300.719\kern0.1em s
	&0.150\kern0.1em s
	&2005
	\\
\hline
\end{tabular}
\end{center}

The case with the second argument bound
still has to be investigated.
The SLDMagic method is no advantage in this case. 
So our Push method would need the same time
as in the \code{tc(X,\kern0.2em Y)} case,
which means 2.171\kern0.2em s for execution.
XSB needs only 0.020\kern0.2em s for execution,
i.e.~is better by a factor of~109!
Since our data loader is quicker,
the factor for total time is less than~2,
but XSB is still better.
Of course,
we are working on improving the SLDMagic method.
But that is a different topic.



\section{The Wine Ontology Benchmark}

\label{secWineOntology}

The wine ontology benchmark
consists of 961~rules with 225~IDB-predicates,
of which all but one are recursive,
and 113~EDB-predicates.
The program is basically one big recursive component.%
\footnote{
Actually,
the wine ontology program does not do what it is supposed to do:
The test query is for californian wines,
but the result contains 
other objects, too.
We checked that XSB produces the same output.
Nevertheless,
the program is a challenging test for a deductive system,
no matter whether it 
is meaningful.
The wine ontology was developed for
the OWL guide \url{http://www.w3.org/TR/owl-guide/}.
It links to a site no longer available,
but the ontology is probably the one available here:
\url{https://www.movesinstitute.org/exi/data/DAML/wines.daml}.
Thanks to Boris Motik,
who referred us to the paper~\cite{HKRT08}:
This introduces an approximate translation from OWL DL TBoxes to Datalog.
Although it is not completely clear yet that this translation was used,
the approximation would explain that the result might contain wrong answers.}
\begin{center}
\begin{tabular}{@{}|l|r|r|r|r|r|@{}}
\hline
System&Load (ms)&Execution (ms)&Total time (ms)&Factor&Memory (KB)\\
\hline
\hline
Push&1\phantom{)}&2\kern0.2em 255&2\kern0.2em 260&1.0&9\kern0.2em 236\\
XSB&106\phantom{)}&8\kern0.2em 548&8\kern0.2em 851&3.9&322\kern0.2em 894\\
YAP&52\phantom{)}&10\kern0.2em 793&10\kern0.2em 899&4.8&334\kern0.2em 761\\
DLV&(60)&---&31\kern0.2em 743&14.0&42\kern0.2em 452\\
\hline
\end{tabular}
\end{center}
This is an example
where the Push method with partial evaluation
as introduced in~\cite{BS15WLP}
``explodes'':
It produces a lot of different specializations of the rules,
of which there are already many in the input program.
Therefore,
we used the version of the push method without partial evaluation
from~\cite{Bra10CPP} here.
Even with that,
the resulting \CPP\ program is large (34\kern0.2em 294~lines).
The detection of duplicates is essential for termination.
We used the hash table here,
probably a bitmap would further improve performance.

Standard implementations of bottom-up evaluation
would iterate all rules in a recursive clique
until one such iteration did not produce a new fact.
In this example,
this would be very inefficient
because basically all rules form a single recursive clique,
but in each iteration only a few rules
actually ``fire''.
In contrast,
the ``Push'' method only looks at rules
which are activated by a new fact for an IDB body literal.
\section{Related Work}

\label{secRelatedWork}

The push method has been studied in~\cite{Bra10CPP},
and compared with ``Pull' and ``Materialization'' methods
of bottom-up evaluation,
but only with artificial examples,
no real data loaded from files,
and no index structures.
The paper~\cite{BS15WLP}
defines a different version of the push method,
which does not create variables for every argument of each IDB predicate,
but for variables occurring in EDB literals in rule bodies.
This helps to reduce or nearly eliminate the copying of values,
at the price of creating several specializations of the same rule
(this reduces runtime, but increases code size, sometimes significantly).
It is basically this version which was used in our benchmarks,
with certain improvements.
Furthermore,
our transformation program can also fall back
to the version of~\cite{Bra10CPP},
if the generated program would otherwise become unacceptably large.
Actually,
both versions can be combined for a single input program.

The idea of immediately using derived facts
to derive more facts is not new in those papers.
For instance,
variants of semi-naive evaluation have been studied
which work in this way~\cite{Sch93,SU99}.
It also seems to be related to the propagation of updates
to materialized views.
However,
the representation of tuples at runtime
and the code structure is different from~\cite{Sch93}
(and this is essential for the reduction of copying values).
The paper~\cite{SU99} 
translates from a temporal Datalog extension to Prolog,
which makes any efficiency comparison dependend
on implementation details of the used Prolog compiler.

In relational databases, the benefits of not materializing intermediate
results have been recognized for a long time; this is known as 
\emph{pipelining}. Pushing data up a pipeline has been studied in 
\cite{Neu11}; 
however, they work with relational algebra expressions rather than
Datalog rules.




\section{Conclusion}

\label{secConclusion}

This is a paper about some experiences gained
while implementing a deductive database system
and testing its performance.
Of course, the system is not yet finished,
and more problems and insights are waiting along the road.
Nevertheless,
the understanding reached so far seems interesting and useful:
\begin{itemize}
\item
A system based on a bottom-up method can compete with
systems based on SLD-resolution with tabling.
Older deductive database systems like Coral~\cite{RSS94,SFH96}
had no chance against XSB~\cite{SSW94}.
Avoiding the copying of data values
and the materialization of derived tuples
seems to be the main reason for the success of the Push method.
\item
Fast data structure implementations,
especially for duplicate elimination,
are very important.
For duplicate tests,
tuples were materialized in ``set'' data structures.
We plan to use nested tables in future											
in order to keep the advantage of reducing the copying of values.
\item
The SLDMagic method proved very useful
for the transitive closure with binding pattern~\code{bf}.
For the binding pattern~\code{fb},
work is still needed.
\item
Some variants in the code
that intuitively seemed important
had no effect on performance.
Probably,
the \code{g++} compiler is clever enough
to optimize the different versions in the same way.
For instance,
we first tried a static procedure which only counted the derived facts,
but did not return them.
When we later used an object with a cursor interface
where one can fetch each result
there was no difference
(although there were many more procedure calls,
and what had been local or static variables before
had to become attributes).
Furthermore,
replacing a \code{switch} and \code{goto}
(our standard translation)
with \code{while}-loops (where possible)
had no important influence on performance
(but improves the readability of the generated code, of course).
\end{itemize}
%
We plan to define an abstract machine											
and translate alternatively into this machine.
Then it would not always be necessary to use a \CPP\ compiler.						
%
%
%

We also work on a parallelized version of our method.
At times																		
where even small PCs have at least four cores,
one should obviously make use of this resource.
The easier parallelization is also one motivation									
for declarative programming.

Of course,
also adding function symbols (term constructors),
negation and aggregation functions are on the agenda,							
plus more exotic things like ordered predicates and declarative output.			

\bibliographystyle{eptcs}

\bibliography{ddb}

\begin{thebibliography}{10}
\providecommand{\bibitemdeclare}[2]{}
\providecommand{\surnamestart}{}
\providecommand{\surnameend}{}
\providecommand{\urlprefix}{Available at }
\providecommand{\url}[1]{\texttt{#1}}
\providecommand{\href}[2]{\texttt{#2}}
\providecommand{\urlalt}[2]{\href{#1}{#2}}
\providecommand{\doi}[1]{doi:\urlalt{http://dx.doi.org/#1}{#1}}
\providecommand{\bibinfo}[2]{#2}

\bibitemdeclare{inproceedings}{Aref15}
\bibitem{Aref15}
\bibinfo{author}{Molham \surnamestart Aref\surnameend}, \bibinfo{author}{Balder
  \surnamestart ten Cate\surnameend}, \bibinfo{author}{Todd~J. \surnamestart
  Green\surnameend}, \bibinfo{author}{Benny \surnamestart
  Kimelfeld\surnameend}, \bibinfo{author}{Dan \surnamestart
  Olteanu\surnameend}, \bibinfo{author}{Emir \surnamestart Pasalic\surnameend},
  \bibinfo{author}{Todd~L. \surnamestart Veldhuizen\surnameend} \&
  \bibinfo{author}{Geoffrey \surnamestart Washburn\surnameend}
  (\bibinfo{year}{2015}): \emph{\bibinfo{title}{Design and Implementation of
  the {LogicBlox} System}}.
\newblock In: {\sl \bibinfo{booktitle}{Proceedings of the 2015 ACM SIGMOD
  International Conference on Management of Data}}, \bibinfo{publisher}{ACM},
  pp. \bibinfo{pages}{1371--1382}, \doi{10.1145/2723372.2742796}.
\newblock
  \urlprefix\url{https://developer.logicblox.com/wp-content/uploads/2016/01/logicblox-sigmod15.pdf}.

\bibitemdeclare{inproceedings}{BMSU86}
\bibitem{BMSU86}
\bibinfo{author}{Francois \surnamestart Bancilhon\surnameend},
  \bibinfo{author}{David \surnamestart Maier\surnameend},
  \bibinfo{author}{Yehoshua \surnamestart Sagiv\surnameend} \&
  \bibinfo{author}{Jeffrey~D. \surnamestart Ullman\surnameend}
  (\bibinfo{year}{1986}): \emph{\bibinfo{title}{Magic Sets and Other Strange
  Ways to Implement Logic Programs}}.
\newblock In: {\sl \bibinfo{booktitle}{Proc.~of the 5th ACM Symp.~on Principles
  of Database Systems (PODS'86)}}, \bibinfo{publisher}{ACM Press}, pp.
  \bibinfo{pages}{1--15}, \doi{10.1145/6012.15399}.

\bibitemdeclare{inproceedings}{BR87}
\bibitem{BR87}
\bibinfo{author}{Catril \surnamestart Beeri\surnameend} \&
  \bibinfo{author}{Raghu \surnamestart Ramakrishnan\surnameend}
  (\bibinfo{year}{1987}): \emph{\bibinfo{title}{On the Power of Magic}}.
\newblock In: {\sl \bibinfo{booktitle}{Proc.~of the Sixth ACM
  SIGACT-SIGMOD-SIGART Symposium on Principles of Database Systems (PODS'87)}},
  \bibinfo{publisher}{ACM}, pp. \bibinfo{pages}{269--284},
  \doi{10.1145/28659.28689}.

\bibitemdeclare{inproceedings}{Bra00}
\bibitem{Bra00}
\bibinfo{author}{Stefan \surnamestart Brass\surnameend} (\bibinfo{year}{2000}):
  \emph{\bibinfo{title}{{SLDMagic} --- The Real Magic (with Applications to Web
  Queries).}}
\newblock In \bibinfo{editor}{W.~\surnamestart Lloyd\surnameend} et~al.,
  editors: {\sl \bibinfo{booktitle}{First International Conference on
  Computational Logic (CL'2000/DOOD'2000)}}, {\sl \bibinfo{series}{LNCS}}
  \bibinfo{volume}{1861}, \bibinfo{publisher}{Springer}, pp.
  \bibinfo{pages}{1063--1077}, \doi{10.1007/3-540-44957-4\_71}.
\newblock
  \urlprefix\url{http://users.informatik.uni-halle.de/~brass/sldmagic/}.

\bibitemdeclare{inproceedings}{Bra10CPP}
\bibitem{Bra10CPP}
\bibinfo{author}{Stefan \surnamestart Brass\surnameend} (\bibinfo{year}{2010}):
  \emph{\bibinfo{title}{Implementation Alternatives for Bottom-Up Evaluation}}.
\newblock In \bibinfo{editor}{Manuel \surnamestart Hermenegildo\surnameend} \&
  \bibinfo{editor}{Torsten \surnamestart Schaub\surnameend}, editors: {\sl
  \bibinfo{booktitle}{Technical Communications of the 26th International
  Conference on Logic Programming (ICLP'10)}}, {\sl \bibinfo{series}{Leibniz
  International Proceedings in Informatics (LIPIcs)}}~\bibinfo{volume}{7},
  \bibinfo{publisher}{Schloss Dagstuhl}, pp. \bibinfo{pages}{44--53},
  \doi{10.4230/LIPIcs.ICLP.2010.44}.
\newblock \urlprefix\url{http://users.informatik.uni-halle.de/~brass/botup/}.

\bibitemdeclare{inproceedings}{Bra12}
\bibitem{Bra12}
\bibinfo{author}{Stefan \surnamestart Brass\surnameend} (\bibinfo{year}{2012}):
  \emph{\bibinfo{title}{Order in Datalog with Applications to Declarative
  Output}}.
\newblock In \bibinfo{editor}{Pablo \surnamestart Barcel{\'o}\surnameend} \&
  \bibinfo{editor}{Reinhard \surnamestart Pichler\surnameend}, editors: {\sl
  \bibinfo{booktitle}{Datalog in Academica and Industry, 2nd Int.~Workshop,
  Datalog~2.0}}, {\sl \bibinfo{series}{LNCS}} \bibinfo{volume}{7494},
  \bibinfo{publisher}{Springer-Verlag}, pp. \bibinfo{pages}{56--67},
  \doi{10.1007/978-3-642-32925-8\_7}.
\newblock \urlprefix\url{http://users.informatik.uni-halle.de/~brass/order/}.

\bibitemdeclare{inproceedings}{BS15WLP}
\bibitem{BS15WLP}
\bibinfo{author}{Stefan \surnamestart Brass\surnameend} \&
  \bibinfo{author}{Heike \surnamestart Stephan\surnameend}
  (\bibinfo{year}{2015}): \emph{\bibinfo{title}{Bottom-Up Evaluation of
  Datalog: Preliminary Report}}.
\newblock In \bibinfo{editor}{Sibylle \surnamestart Schwarz\surnameend} \&
  \bibinfo{editor}{Steffen \surnamestart H{\"o}lldobler\surnameend}, editors:
  {\sl \bibinfo{booktitle}{29th Workshop on (Constraint) Logic Programming (WLP
  2015)}}, \bibinfo{publisher}{HTWK Leipzig}, pp. \bibinfo{pages}{21--35}.
\newblock \urlprefix\url{http://www.imn.htwk-leipzig.de/WLP2015/}.

\bibitemdeclare{article}{CW96SLG}
\bibitem{CW96SLG}
\bibinfo{author}{Weidong \surnamestart Chen\surnameend} \&
  \bibinfo{author}{David~S. \surnamestart Warren\surnameend}
  (\bibinfo{year}{1996}): \emph{\bibinfo{title}{Tabled Evaluation with Delaying
  for General Logic Programs}}.
\newblock {\sl \bibinfo{journal}{Journal of the ACM}}
  \bibinfo{volume}{43}(\bibinfo{number}{1}), pp. \bibinfo{pages}{20--74},
  \doi{10.1145/227595.227597}.

\bibitemdeclare{inproceedings}{Costa99}
\bibitem{Costa99}
\bibinfo{author}{V{\'\i}tor~Santos \surnamestart Costa\surnameend}
  (\bibinfo{year}{1999}): \emph{\bibinfo{title}{Optimizing Bytecode Emulation
  for Prolog}}.
\newblock In \bibinfo{editor}{Gopalan \surnamestart Nadathur\surnameend},
  editor: {\sl \bibinfo{booktitle}{Principles and Practice of Declarative
  Programming, International Conference PPDP'99}}, {\sl \bibinfo{series}{LNCS}}
  \bibinfo{volume}{1702}, \bibinfo{publisher}{Springer}, pp.
  \bibinfo{pages}{261--277}, \doi{10.1007/10704567\_16}.

\bibitemdeclare{article}{CRD12}
\bibitem{CRD12}
\bibinfo{author}{V\'{\i}tor~Santos \surnamestart Costa\surnameend},
  \bibinfo{author}{Ricardo \surnamestart Rocha\surnameend} \&
  \bibinfo{author}{Lu\'{\i}s \surnamestart Damas\surnameend}
  (\bibinfo{year}{2012}): \emph{\bibinfo{title}{The {YAP} Prolog System}}.
\newblock {\sl \bibinfo{journal}{Theory and Practice of Logic Programming}}
  \bibinfo{volume}{12}(\bibinfo{number}{1--2}), pp. \bibinfo{pages}{5--34},
  \doi{10.1017/S1471068411000512}.
\newblock
  \urlprefix\url{https://www.dcc.fc.up.pt/~ricroc/homepage/publications/2012-TPLP.pdf}.

\bibitemdeclare{inproceedings}{HKRT08}
\bibitem{HKRT08}
\bibinfo{author}{Pascal \surnamestart Hitzler\surnameend},
  \bibinfo{author}{Markus \surnamestart Kr{\"o}tzsch\surnameend},
  \bibinfo{author}{Sebastian \surnamestart Rudolph\surnameend} \&
  \bibinfo{author}{Tuvshintur \surnamestart Tserendorj\surnameend}
  (\bibinfo{year}{2008}): \emph{\bibinfo{title}{Approximate {OWL} Instance
  Retrieval with {SCREECH}}}.
\newblock In \bibinfo{editor}{Anthony~G. \surnamestart Cohn\surnameend},
  \bibinfo{editor}{David~C. \surnamestart Hogg\surnameend},
  \bibinfo{editor}{M{\"o}ller. \surnamestart Ralf\surnameend} \&
  \bibinfo{editor}{Bernd \surnamestart Neumann\surnameend}, editors: {\sl
  \bibinfo{booktitle}{Logic and Probability for Scene Interpretation}}, {\sl
  \bibinfo{series}{Dagstuhl Seminar Proceedings}} \bibinfo{volume}{08091},
  \bibinfo{publisher}{Schloss Dagstuhl - Leibniz-Zentrum fuer Informatik,
  Germany}, \bibinfo{address}{Dagstuhl, Germany}, pp. \bibinfo{pages}{1--8}.
\newblock \urlprefix\url{http://drops.dagstuhl.de/opus/volltexte/2008/1615}.

\bibitemdeclare{inproceedings}{LKN13}
\bibitem{LKN13}
\bibinfo{author}{Viktor \surnamestart Leis\surnameend}, \bibinfo{author}{Alfons
  \surnamestart Kemper\surnameend} \& \bibinfo{author}{Thomas \surnamestart
  Neumann\surnameend} (\bibinfo{year}{1997}): \emph{\bibinfo{title}{The
  Adaptive Radix Tree: {ARTful} Indexing for Main-Memory Databases}}.
\newblock In: {\sl \bibinfo{booktitle}{Proc.~of the 2013 IEEE International
  Conference on Data Engineering (ICDE'2013)}}, \bibinfo{publisher}{IEEE
  Computer Society}, pp. \bibinfo{pages}{38--49},
  \doi{10.1109/ICDE.2013.6544812}.
\newblock
  \urlprefix\url{http://www3.informatik.tu-muenchen.de/~leis/papers/ART.pdf}.

\bibitemdeclare{article}{LPFEGPS06}
\bibitem{LPFEGPS06}
\bibinfo{author}{Nicola \surnamestart Leone\surnameend},
  \bibinfo{author}{Gerald \surnamestart Pfeifer\surnameend},
  \bibinfo{author}{Wolfgang \surnamestart Faber\surnameend},
  \bibinfo{author}{Thomas \surnamestart Eiter\surnameend},
  \bibinfo{author}{Georg \surnamestart Gottlob\surnameend},
  \bibinfo{author}{Simona \surnamestart Perri\surnameend} \&
  \bibinfo{author}{Francesco \surnamestart Scarcello\surnameend}
  (\bibinfo{year}{2006}): \emph{\bibinfo{title}{The DLV system for knowledge
  representation and reasoning}}.
\newblock {\sl \bibinfo{journal}{ACM Trans. Comput. Logic}}
  \bibinfo{volume}{7}(\bibinfo{number}{3}), pp. \bibinfo{pages}{499--562},
  \doi{10.1145/1149114.1149117}.
\newblock \urlprefix\url{https://arxiv.org/pdf/cs/0211004}.

\bibitemdeclare{inproceedings}{LFWK09}
\bibitem{LFWK09}
\bibinfo{author}{Senlin \surnamestart Liang\surnameend}, \bibinfo{author}{Paul
  \surnamestart Fodor\surnameend}, \bibinfo{author}{Hui \surnamestart
  Wan\surnameend} \& \bibinfo{author}{Michael \surnamestart Kifer\surnameend}
  (\bibinfo{year}{2009}): \emph{\bibinfo{title}{{OpenRuleBench}: {An} Analysis
  of the Performance of Rule Engines}}.
\newblock In: {\sl \bibinfo{booktitle}{Proceedings of the 18th International
  Conference on World Wide Web (WWW'09)}}, \bibinfo{publisher}{ACM}, pp.
  \bibinfo{pages}{601--610}, \doi{10.1145/1526709.1526790}.
\newblock \urlprefix\url{http://rulebench.projects.semwebcentral.org/}.

\bibitemdeclare{article}{Neu11}
\bibitem{Neu11}
\bibinfo{author}{Thomas \surnamestart Neumann\surnameend}
  (\bibinfo{year}{2011}): \emph{\bibinfo{title}{Efficiently Compiling Efficient
  Query Plans for Modern Hardware}}.
\newblock {\sl \bibinfo{journal}{Proceedings of the VLDB Endowment}}
  \bibinfo{volume}{4}(\bibinfo{number}{9}), pp. \bibinfo{pages}{539--550},
  \doi{10.14778/2002938.2002940}.
\newblock \urlprefix\url{http://www.vldb.org/pvldb/vol4/p539-neumann.pdf}.

\bibitemdeclare{article}{RSS94}
\bibitem{RSS94}
\bibinfo{author}{Raghu \surnamestart Ramakrishnan\surnameend},
  \bibinfo{author}{Divesh \surnamestart Srivastava\surnameend} \&
  \bibinfo{author}{S.~\surnamestart Sudarshan\surnameend}
  (\bibinfo{year}{1994}): \emph{\bibinfo{title}{Rule Ordering in Bottom-Up
  Fixpoint Evaluation of Logic Programs}}.
\newblock {\sl \bibinfo{journal}{IEEE Transactions on Knowledge and Data
  Engineering}} \bibinfo{volume}{6}(\bibinfo{number}{4}), pp.
  \bibinfo{pages}{501--517}, \doi{10.1109/69.298169}.
\newblock
  \urlprefix\url{https://www.cse.iitb.ac.in/~sudarsha/Pubs-dir/ruleordering-tkde.pdf}.

\bibitemdeclare{inproceedings}{Ros91}
\bibitem{Ros91}
\bibinfo{author}{Kenneth~A. \surnamestart Ross\surnameend}
  (\bibinfo{year}{1991}): \emph{\bibinfo{title}{Modular Acyclicity and Tail
  Recursion in Logic Programs}}.
\newblock In: {\sl \bibinfo{booktitle}{Proc.~of the Tenth ACM
  SIGACT-SIGMOD-SIGART Symp.~on Princ.~of Database Systems (PODS'91)}}, pp.
  \bibinfo{pages}{92--101}, \doi{10.1145/113413.113422}.

\bibitemdeclare{inproceedings}{SSW94}
\bibitem{SSW94}
\bibinfo{author}{Konstantinos \surnamestart Sagonas\surnameend},
  \bibinfo{author}{Terrance \surnamestart Swift\surnameend} \&
  \bibinfo{author}{David~S. \surnamestart Warren\surnameend}
  (\bibinfo{year}{1994}): \emph{\bibinfo{title}{{XSB} as an Efficient Deductive
  Database Engine}}.
\newblock In \bibinfo{editor}{Richard~T. \surnamestart Snodgrass\surnameend} \&
  \bibinfo{editor}{Marianne \surnamestart Winslett\surnameend}, editors: {\sl
  \bibinfo{booktitle}{Proc.~of the 1994 ACM SIGMOD Int.~Conf.~on Management of
  Data (SIGMOD'94)}}, pp. \bibinfo{pages}{442--453},
  \doi{10.1145/191843.191927}.
\newblock \urlprefix\url{http://user.it.uu.se/~kostis/Papers/xsbddb.html}.

\bibitemdeclare{phdthesis}{Sch93}
\bibitem{Sch93}
\bibinfo{author}{Heribert \surnamestart Sch{\"u}tz\surnameend}
  (\bibinfo{year}{1993}): \emph{\bibinfo{title}{{T}upelweise
  {B}ottom-up-{A}uswertung von {L}ogikprogrammen (Tuple-wise bottom-up
  evaluation of logic programs)}}.
\newblock Ph.D. thesis, \bibinfo{school}{TU M{\"u}nchen}.

\bibitemdeclare{techreport}{SFH96}
\bibitem{SFH96}
\bibinfo{author}{Praveen \surnamestart Seshadri\surnameend},
  \bibinfo{author}{Shaun \surnamestart Flisakowski\surnameend} \&
  \bibinfo{author}{Seymour \surnamestart Hersh\surnameend}
  (\bibinfo{year}{1996}): \emph{\bibinfo{title}{CORAL: The Inside Story.
  {Shocking} Hacks Revealed}}.
\newblock \bibinfo{type}{Technical Report}, \bibinfo{institution}{Department of
  Computer Sciences, The University of Wisconsin-Madison}.
\newblock \urlprefix\url{http://ftp.cs.wisc.edu/coral/doc/Inside.ps}.

\bibitemdeclare{inproceedings}{SU99}
\bibitem{SU99}
\bibinfo{author}{Donald~A. \surnamestart Smith\surnameend} \&
  \bibinfo{author}{Mark \surnamestart Utting\surnameend}
  (\bibinfo{year}{1999}): \emph{\bibinfo{title}{Pseudo-Naive Evaluation}}.
\newblock In: {\sl \bibinfo{booktitle}{Australasian Database Conference}}, pp.
  \bibinfo{pages}{211--223}.
\newblock
  \urlprefix\url{http://citeseerx.ist.psu.edu/viewdoc/summary?doi=10.1.1.177.5047}.

\bibitemdeclare{article}{Swi99}
\bibitem{Swi99}
\bibinfo{author}{Terrance \surnamestart Swift\surnameend}
  (\bibinfo{year}{1999}): \emph{\bibinfo{title}{Tabling for non-monotonic
  programming}}.
\newblock {\sl \bibinfo{journal}{Annals of Mathematics and Artificial
  Intelligence}} \bibinfo{volume}{25}, pp. \bibinfo{pages}{201--240},
  \doi{10.1023/A:1018990308362}.
\newblock
  \urlprefix\url{http://www3.cs.stonybrook.edu/~tswift/webpapers/amai-99.ps}.

\end{thebibliography}

\end{document}